# A Negotiation-based Right-of-way Assignment Strategy to Ensure Traffic Safety and Efficiency in Lane Changes

Can Zhao [1], Zhiheng Li [2], Li Li [*1], Xiangbin Wu [3], Fei-Yue Wang [4]

[1] Department of Automation, Tsinghua University, Beijing, China
[2] Tsinghua Shenzhen International Graduate School, Tsinghua University, Shenzhen, China
[3] Intel Labs China, Intel Corporation, Beijing, China
[4] The State Key Laboratory for Management and Control of Complex Systems, Institute of Automation, Chinese Academy of Sciences, Beijing, China
[*] li-li@tsinghua.edu.cn

**Abstract:** It is widely acknowledged that verifying the safety of autonomous driving strategies requires a substantial body of simulation testing and road testing. In recent years, the formal safety methods represented by Responsibility-Sensitive Safety (RSS) have encouraged low-cost autonomous driving safety research, benefitting from its accurate assessment of safety and clear division of responsibilities. However, how to maintain traffic efficiency while ensuring safety remains a challenge. To address this problem, this paper proposes a formulized negotiation-based lane-changing strategy that makes a trade-off between safety and efficiency. Both theoretical analysis and numerical experimental results shows that compared to RSS, our strategy can noticeably improve the success rate of changing lanes on the premise of safety.

## 1. Introduction

Safety is the first criterion for autonomous vehicles (AVs). The original intention of this technology is to bring more certainty and reliability to road traffic and reduce the damage caused by accidents, i.e., fewer crashes and injuries. Generally, verifying the safety of autonomous driving strategies requires a substantial body of simulation testing and road testing. However, how to ensure that the testing environment faithfully describes actual road conditions (including many corner cases) remains difficulty.

In recent years, researchers have tried to verify the safety of AVs by formulating a set of standardized maneuvers. The proposal of formulized verification methods has encouraged low-cost autonomous driving safety research [1]. Specifically, formulized safety methods can determine when to make dedicated actions within the given scenarios and verify the feasibility and correctness based on uncomplicated mathematical calculations.

For example, based on the principle of "absolute safety," the Responsibility-Sensitive Safety (RSS) strategy proposed by Mobileye is a typical formulized safety model. Mobileye positions RSS as an instructive and scalable autonomous driving strategy system, aiming to call on the autonomous driving industry (technology providers, researchers and lawmaking agencies) to standardize driving rules and responsibilities [2]. To date, RSS has been tested in 37 typical accident scenarios covering 99.4% of National Highway Safety Administration (NHTSA) accident scenario data, and the results have shown that its security has reached a sound available state [3]. Whether these formulized safety methods can be applied in more complex driving scenarios has attracted increasing attention.

This paper focuses on the performance and improvements of formulized strategies in lane-changing scenarios [4-6]. According to data from the NHTSA, traffic accidents caused by lane changes account for up to 27% of all [7], which has become an enormous challenge for both human drivers and AVs.

In this paper, we propose a communication-based formulized lane-changing strategy that addresses the assignment of right-of-way to ensure traffic safety and efficiency simultaneously. Based on our previous study [8], it is apparent that via correct communications, the whole decision problem can be easily decomposed into several much simpler subproblems based on the transferring time points of right-of-way. We call this method a situation-aware strategy for specified tasks within particular driving scenarios [9-11]. Based on the original RSS strategy, this paper divides the lane-changing maneuver process into three stages clearly, and formulates the safety conditions in each time interval during each stage. The simulation results indicate that our strategy noticeably improves the success rate of changing lanes on the premise of safety.

The contributions can be summarized as follows:

(1) This work defines the concept and basic principles of the right-of-way and verifies its feasibility in lane-changing scenarios with a rule-based formulized method. Furthermore, this method can identify the parties responsible for potential accidents quickly and accurately.

(2) This work coordinates the driving intentions of all participants by introducing an interactive negotiation mechanism. In addition to lane-changing scenarios, this idea also has prospects for application to other traffic scenarios.

(3) This work not only embodies the requirements of traffic efficiency but also gives enough consideration to driving safety. We take both multivehicle lane-changing cases and negotiation failure cases into account.

To better present our findings, the rest of this paper is arranged as follows. *Section 2* systematically reviews studies on formal lane-changing strategies and highlights their lack of attention to negotiation. *Section 3* defines the concept and principles of the right-of-way, and introduces the scenarios we focused on. *Section 4* provides numerical experimental results to verify the superiority of our new strategy over RSS on traffic efficiency. *Section 5* provides an additional discussion on multivehicle lane-changing cases and negotiation failure cases and performs simulations to verify the security of the new strategy. Finally, *Section 6* concludes the paper and briefly describes the next steps of our research.

## 2. Literature Review

### 2.1 Correlational Research

The decision-making process answers the question "under what circumstances should an AV make a lane-changing decision." Most of this kind of research originates from the modeling of human behaviors [12-15]. The MITSIM model [12] was proposed in 2002 and has since been amended several times. Most studies based on it [16-17] divided the lane-changing process into three critical stages: i) determine whether there is a need to change; ii) judge the acceptance gaps, and iii) make the lane change. The lane change decision can also be regarded as a game process between merging vehicles and straight-going vehicles [18-20], and traffic rules are generally considered implicit in the respective decision models. Other typical methods include logical reasons [21] and data-driven ensemble learning [22]. Nevertheless, these models did not clearly define how to avoid collisions altogether [23].

The safety verification process answers the question "how to ensure that the lane-changing decision is safe." Methods such as *Multi-Lane Spatial Logic* (MLSL), *Inevitable Collision States* (ICS), and *Reachability Analysis* (RA) are widely used in this field. In Reference [24], researchers have used MLSL to verify the safety of lane changes. This method assesses safety by monitoring whether the space occupied by the vehicles (including the vehicle itself and the reserved space) intersects. ICS packages all the unsafe states of AVs into a set and considers all driving behaviors avoiding them as safe states [25-26]. RA focuses on the intersection of the feasible trajectory from all traffic participants [27], which provides a testability criterion for a driving strategy. However, these methods rely on an accurate perception system and an enormous amount of computation, leading to poor performance in applicability and timeliness.

Reference [6] verifies the feasibility of lane changes by formalizing traffic rules and maneuvering processes, coinciding with the idea of RSS. In addition to the reduction in computation, another advantage of this kind of method is the explicit division of collision responsibility [1]. Plainly speaking, if all participants comply with the driving principles of model making, traffic accidents will be avoided entirely. When an accident occurs, the first rule violator shall bear the whole responsibility for all consequences.

### 2.2 Weaknesses and Improvements

However, we have noted room for improvement in the lane-changing strategies of both RSS and the work in [6].

First, both strategies recognize that the essence of safe driving is assigning right-of-way properly for all traffic participants, but they do not clearly define the concept of right-of-way and the principle of assignment. Obviously, if every vehicle is connected and autonomous (CAV), we can utilize *explicit communication* techniques (e.g., V2V) [28-30] to arrange their right-of-way effortlessly. While in the foreseeable future, traditional human-driven vehicles and AVs will coexist for a long time. Therefore, the assignment of right-of-way under mixed traffic flows must be compatible with the existing rules and habits of human drivers. The most important principle among is the *First Come First Served* (FCFS) [31-33].

Second, both strategies consider the vehicle simply as an isolated individual and ignore the interactions among vehicles. However, the efficient completion of the right-of-way assignment process can never be divorced from communication and negotiation [34]. In road traffic, the exchange of information is ubiquitous. In human driving, there has always existed primitive information transmission (e.g., horns, gestures, lights, and speed changes) not relying on specific communication equipment, which we call *implicit communication* [38]. Restricted by this information transfer channel, the communication efficiency may be low, but it is still essential for driving safety [35-37]. For example, in lane-changing scenarios, the acceleration or deceleration of the following vehicle in the target lane may reflect reverse intentions. In recent years, identifying the human driving intention from implicit communication has become a research topic of interest [39-41]. Although this paper does not focus on intention recognition, we believe that the efficiency and success rate of negotiation will continue to improve as research in this area progresses.

Moreover, due to a lack of communication, existing defensive driving strategies tend to behave quite conservatively, which may indirectly affect traffic efficiency [38]. For example, [8] demonstrated that safety and efficiency can be better balanced in vehicle-following scenarios by slightly modifying the original RSS strategy. Similarly, the lane-changing opportunities calculated by the RSS also tend to be overly conservative, which makes lane changes a luxury in many cases. As a comparison, human drivers always use turn signals to give early warnings to the following vehicle, promoting awareness and position adjustment. Thus, we are interested in studying whether we can improve the lane-changing performance of AVs by introducing a negotiation process [38].

## 3. Problem Presentation

### 3.1 Assumptions

The major assumptions involved in the following sections are listed below:

A1) We assume that the lane-changing maneuvers implemented in this paper belong to *discretionary lane changes* (DLCs), and *mandatory lane changes* (MLCs) can be addressed via a similar approach [5, 42]. Unlike MLCs (e.g., entering and exiting ramp junctions [43-44], multilane merging, etc.), DLCs do not have strict requirements for lane-changing times and locations; they occur to earn better driving benefits (e.g., visibility, expected speed, etc.).

A2) Only one-way scenarios are considered. Meanwhile, $V_{\text{ego}}$ is in a safe state and has determined the target lane at the beginning.

A3) $V_{\text{ego}}$ can acquire the state (position, speed, and acceleration) of surrounding vehicles via sensors. In addition,

we assume that the outside size and the dynamic performance of all studied vehicles are the same.

A4) At the same moment, the full width of a lane should be occupied by only one vehicle in the lane, which is a common conservative setting in lane-changing decision-making studies [6, 24]. Of course, further divisions of the lateral area are necessary when generating specific maneuvering schemes [45-46].

A5) We assume $V_{ego}$ is the only merging vehicle during the entire lane-changing process.

A6) We assume that all vehicles are rational, meaning that they are all aware of the same right-of-way principles and willing to cooperate in negotiations.

Among these assumptions, A1-A4 simplify the scenarios and eliminate noncritical information, and A5-A6 reduce the calculation of collision conditions. To assure the integrity and scientific nature of this work, we specifically discuss the failure cases of A5 and A6 in Section 5.

### 3.2 The Definition and Principles of Right-of-way

The right-of-way is a central concept in traditional traffic regulations, but only has limited guidance for autonomous driving strategy generation. First, this is because the right-of-way concept is polysemous in different countries [47-49] and lacks uniform standards. As a result, many safety driving frameworks, represented by RSS, were originally introduced to facilitate cross-industry discussions and consensus among industry organizations, manufacturers, and regulators [2]. Second, the primary purpose of traditional traffic regulations is to determine and assign responsibility after a traffic violation rather than to provide predictive driving guidance. Therefore, the rules of right-of-way transfer are not clearly defined, and only liability decisions such as "the merging vehicle is responsible for not touching the vehicle straight going " and "the following vehicle is responsible for a rear-end accident" are made. For human drivers, such "posteriori" traffic regulations can provide an implicit behavioral boundary; however, for rule-based AVs with higher safety requirements, a more specific, calculable, and interpretable right-of-way model is required.

RSS highlights the role of the right-of-way playing in road safety. However, it did not explicitly answer two crucial questions: "what is the right-of-way" and "when to transfer the right-of-way." A thorough rethinking of human driving behaviors shows that if there are no more specific criteria (e.g., traffic rules or threat of danger), most human drivers will determine the right-of-way ownership of a certain conflict area based on who can occupy it faster and easier. This is known as the FCFS principle, which has been proven also helpful for AVs in improving driving safety and efficiency because of the low fuzziness and clear calculation process [28].

Therefore, referring to the prevailing academic definition [50-51], we finally define the right-of-way as *the preferential right to occupy/use a certain temporal-spatial area* [52]. Furthermore, based on FCFS and safety requirements, we summarize the following four basic principles of the right-of-way that should be respected:

1) The existence of the right-of-way is relative. Only when there is a potential conflict between two traffic participants does the right of way assignment arise.

2) Within the right-of-way area, there is only one owner at every moment. The nonowner shall take the initiative to keep out of this area unless permission is obtained from the owner.

3) When there is a conflict between owner and nonowner in a certain area, nonowner are responsible for potential accidents.

4) Depending on whether the owner can transfer, the right-of-way areas can be classified as *forbidden areas* and *negotiable areas*. The former is inviolable, and the latter can be transferred to another owner after negotiation.

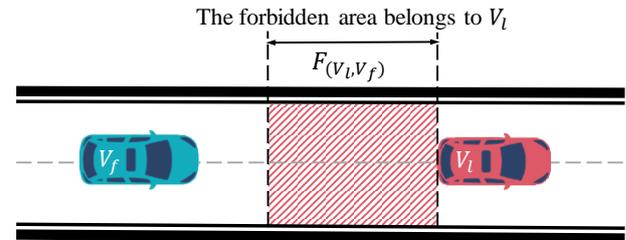

*Fig. 1.* *The assignment of right-of-way in vehicle-following scenarios*

Taking the vehicle-following scenario shown in Fig. 1 as an example, $V_l$ (the leading vehicle) naturally possesses the right-of-way of the certain area behind it, which means that $V_f$ (the following vehicle) has responsibility for keeping away from this area to ensure safety. According to our own work [8], the length of this forbidden area $F_{(V_l,V_f)}$ can be calculated based on the speed of the two vehicles:

$$F_{(V_l,V_f)} = \left[ v_f \rho + \frac{v_f^2}{2a_{f,brake}} - \frac{v_l^2}{2a_{max,brake}} \right]^+ \quad (1\text{-}1)$$

where

$$a_{f,brake} = a_{min,brake} + \frac{v_f}{v_{max}}\left(a_{max,brake} - a_{min,brake}\right) \quad (1\text{-}2)$$

where $v_f$ denotes the speed of $V_f$, $v_l$ denotes the speed of $V_l$, $v_{max}$ denotes the maximum speed limit of the lane, and $\rho$ denotes the reaction time lag. $F_{(V_l,V_f)}$ can be interpreted as follows: *if the leading vehicle suddenly brakes with maximum deceleration, this distance accommodates the following vehicle to safely stop with* $a_{f,brake}$ *after the reaction time.*

### 3.3 Scenario Presentation

The nomenclatures used in this paper are given in Table 1.

We model the DLC scenario in Fig. 2 and denote the merging autonomous vehicle as $V_{ego}$, which aims to switch from $L_1$ (the original lane) to $L_2$ (the target lane). $V_{ego}$ may interact with the surrounding four vehicles on the right of way, including $V_{l1}$ (the leading vehicle of $L_1$), $V_{f1}$ (the following vehicle of $L_1$), $V_{l2}$ (the leading vehicle of $L_2$), and $V_{f2}$ (the following vehicle of $L_2$).

*Table 1. The Nomenclature List*

| Symbol | Definition |
|---|---|
| $l_v$ | The length of a vehicle |
| $F_{(i,j)}$ | The length of the forbidden area between $i$ and $j$; the right-of-way belongs to $i$ |
| $N_{(i,j)}$ | The length of the negotiable area between $i$ and $j$; the right-of-way belongs to $i$ |
| $d_{(i,j)}$ | The longitudinal distance between $i$ and $j$ |
| $\rho$ | The reaction time lag of AV |
| $\rho_{human}$ | The reaction time lag of human driver |
| $v_i$ | The longitudinal speed of $V_i$ |
| $a_i$ | The longitudinal acceleration of $V_i$ |
| $a_{max,brake}$ | The maximum deceleration |
| $a_{max,accel}$ | The maximum acceleration |
| $v_{max}$ | The maximum speed limit of the lane |
| $\lambda$ | The traffic flow of the lane |
| $h$ | The headway of the lane |
| $\mu$ | the log-normal location parameter |
| $\sigma$ | the log-normal scale parameter |

Similar to the vehicle-following scenario, $V_{l1}$ and $V_{l2}$ have a natural advantage over $V_{ego}$, so it needs to avoid the forbidden area behind them.

According to the FCFS principle, $V_{f2}$ have the right-of-way for the certain area ahead. The area that poses severe threats to safety should became the forbidden area of $V_{f2}$, and the area that affects driving interests should became the negotiable area.

In Fig. 2, we refer to the direction parallel to the lane as *longitudinal* and the direction perpendicular to it as *lateral*. Given that the considerable speed difference in the two directions, it is sensible to regard a lane-changing maneuver as two independent decoupled uniform motions.

### 3.4 The Human Strategy and the RSS Strategy

Obviously, the ownership of the right-of-way does not need to be updated at every time interval. Instead, a vehicle may hold the right-of-way of a certain area for a long time until another vehicle takes over after obtaining its permission. This property helps us decompose the negotiation and decision process into stages to reduce the complexity of the calculation.

To exhibit the DLCs process visually, we take the human strategy and the original RSS strategy as comparisons, as shown in Fig. 3. The key differences are highlighted in bold.

Experienced human drivers always consider multivehicle right-of-way assignment problems as multiple one-to-one right-of-way assignment problems and solve them in turn. Some researchers divide the human DLCs process into three crucial stages—the *longitudinal spacing adjustment stage*, the *negotiation stage*, and the *action stage* [18]. In the first stage, $V_{ego}$ adjusts the relative position to find a good merging opportunity. In second stage, the human driver will implement courtesy merging under normal conditions, meaning to notify the surrounding vehicles and request permission. However, such estimates of the acceptable gap are subjective and rough, which can trigger disastrous consequences. Moreover, some aggressive drivers may engage in forced merging without negotiation, even in a small headway. This risky behavior is believed to be one of the culprits of accidents.

Conversely, the RSS strategy formulates all determination rules to ensure strict collision-free conditions. This strategy requires $V_{ego}$ to change lanes only when the change not cause any impact on surrounding vehicles, but it does not specify how to make this determination. Furthermore, because of absolute confidence in the safety of lane changing, RSS omits the negotiation process and believes that other vehicle will yield in a timely and initiative fashion. However, this strategic omission introduces several problems. First, to find the right opportunity to change lanes, the expected time for position adjustment will inevitably be extended (proved in *Section 5*). Second, this strategy requires vehicles to yield unconditionally when encounter a lane-changing request. This is of course for safety, but in some cases, it will conflict with driving self-interests, such as being blocked multiple times continuously. Third, the surrounding vehicles can only perceive the intention to change lanes after the merging maneuver begins. In the long run, this may lead to potential risks, such as multiple vehicles merging into the same lane simultaneously.

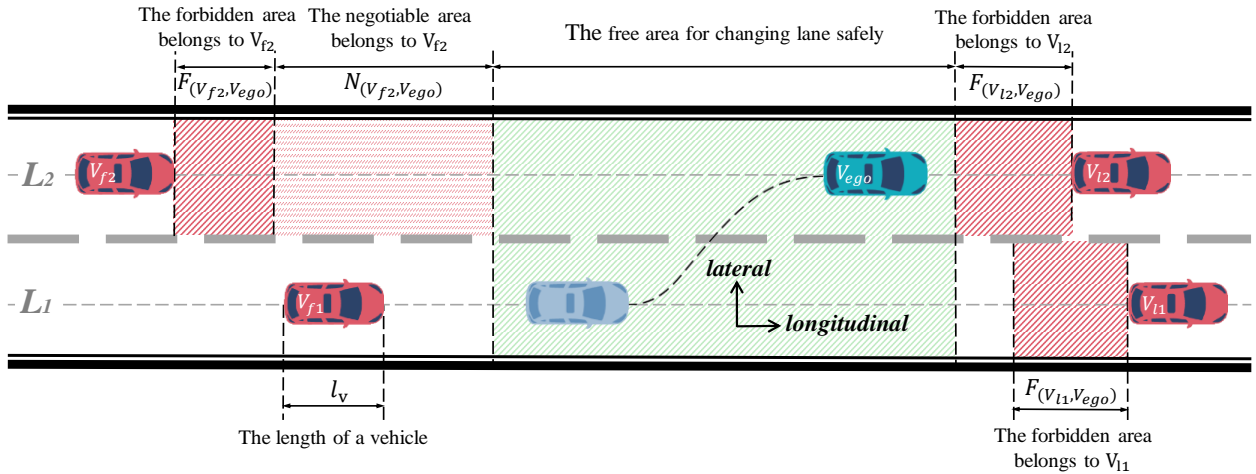

*Fig. 2. An illustration of DLC scenarios.*

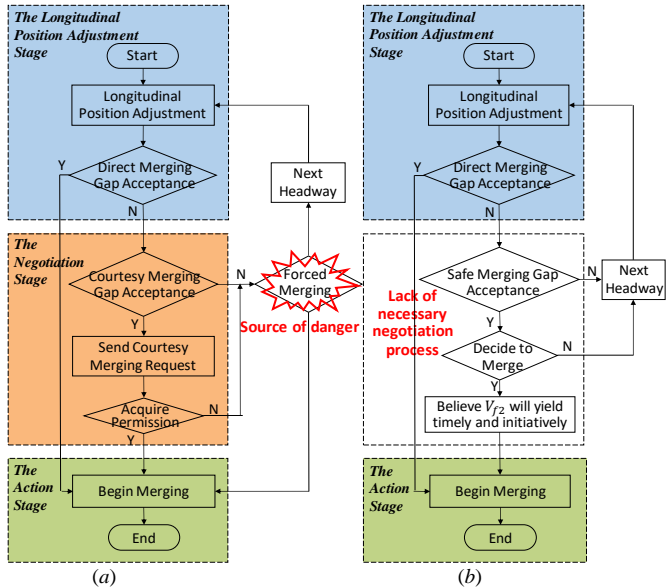

**Fig. 3.** *The flow charts of two existing DLC strategies: (a) the human strategy* [19]; *(b) the RSS strategy* [2]

### 3.5 Three Stages of the New Strategy

We aim to combine the superiorities of the three-stage model and collision avoidance conditions. An ideal lane-changing strategy should be sequential, specific, and easy to calculate. Under ideal conditions, all traffic participants should be allowed to strive for their reasonable driving interests.

Modeled after human driving, the whole merging process is divided into three stages as Fig. 4. We list the state transition conditions of the strategy in this section, and more calculation details are provided in *Section 4*.

*3.5.1 The Longitudinal Position Adjustment Stage.* In this stage, $V_{\text{ego}}$ should observe the state of surrounding vehicles and adjust its position. If it can change lanes without invading the forbidden area of $V_{l1}$ and $V_{l2}$, as well as the right-of-way area of $V_{f2}$, we call this situation *direct merging gap acceptance*. Equation (2) of Fig. 4 can be expressed as:

$$\begin{cases} d_{(V_{ego}, V_{l1})} \geq F_{(V_{l1}, V_{ego})} \\ d_{(V_{ego}, V_{l2})} \geq F_{(V_{l2}, V_{ego})} \\ d_{(V_{ego}, V_{f2})} \geq F_{(V_{f2}, V_{ego})} + N_{(V_{f2}, V_{ego})} \\ a_{f2} \leq 0,\ a_{l1} \geq 0,\ a_{l2} \geq 0 \end{cases} \quad (2)$$

where the variable $d_{(i,j)}$ denotes the longitudinal distance between $i$ and $j$. $F_{(i,j)}$ denotes the length of the *forbidden area* between $i$ (the owner) and $j$ (the nonowner). Similarly, the variable $N_{(i,j)}$ represents the length of the *negotiation area*.

If $V_{\text{ego}}$ will only enter the negotiable area, we call this situation *negotiated merging gap acceptance*. Equation (3) of Fig. 3 can be expressed as:

$$\begin{cases} d_{(V_{ego}, V_{l1})} \geq F_{(V_{l1}, V_{ego})} \\ d_{(V_{ego}, V_{l2})} \geq F_{(V_{l2}, V_{ego})} \\ F_{(V_{f2}, V_{ego})} \leq d_{(V_{ego}, V_{f2})} \leq F_{(V_{f2}, V_{ego})} + N_{(V_{f2}, V_{ego})} \\ a_{f2} \leq 0,\ a_{l1} \geq 0,\ a_{l2} \geq 0 \end{cases} \quad (3)$$

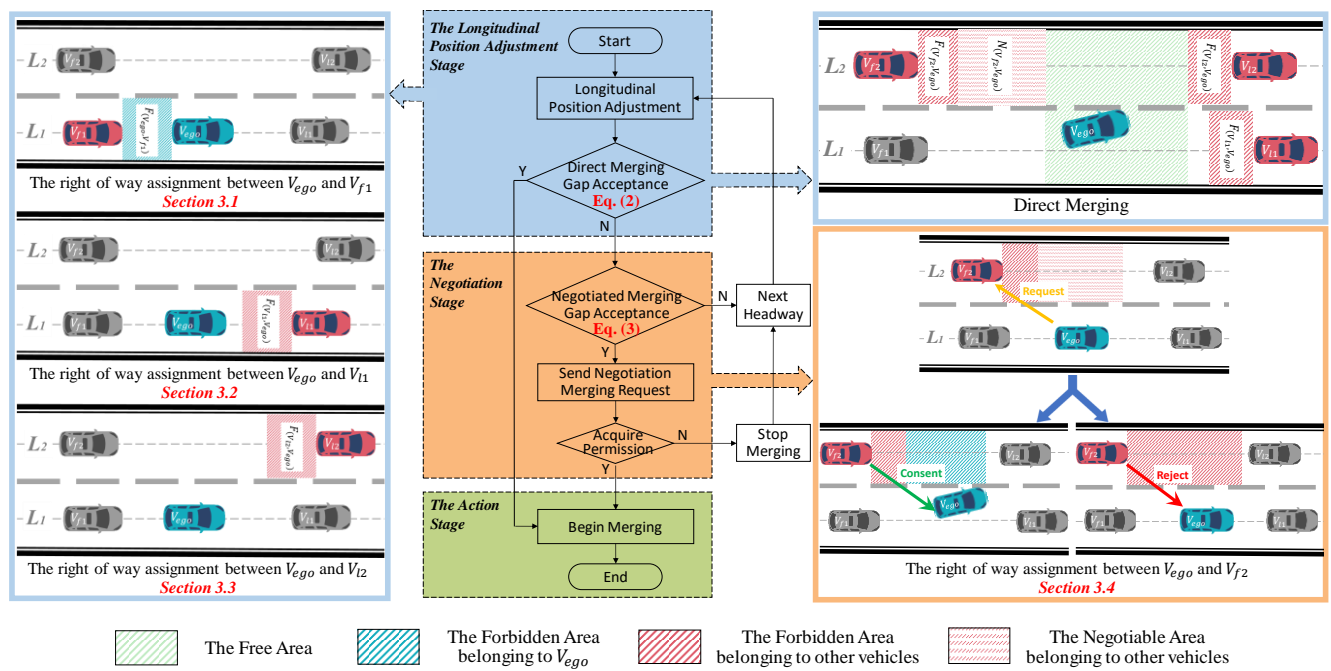

**Fig. 4.** *The flow chart and stages of the new DLCs strategy.*

### 3.5.2 The Right-of-way Negotiation Stage.
Obviously, the negotiation process only occurs between $V_{ego}$ and $V_{f2}$. After $V_{ego}$ has generated the intention to enter the right-of-way area of $V_{f2}$, the issuance of the merge request is necessary. This signal may be delivered by explicit communication or implicit communication (e.g., the turn signal). To ensure that $V_{f2}$ can have sufficient time to respond after receiving the request, we set the feedback waiting time as 3 s after reviewing the literature [54-55]. After this period, there are three possible results for $V_{f2}$:

- *Reject*. If the request is explicitly rejected by $V_{f2}$ (through V2V, acceleration, horn blaring, etc.), $V_{ego}$ should immediately consider the negotiable area as the forbidden area and wait for the next appropriate headway.
- *Consent*. If $V_{f2}$ decelerates or makes no response to acquiescence, the right-of-way ownership of *the negotiation area* will transfer to $V_{ego}$.
- *Failure*. Because of the external similarity between *acquiescence* and *negotiation failure*, the potential risks caused by misjudgments cannot be ignored. Hence, the supplementary discussion of *Section 5* focuses on the safety issues of these exceptional cases.

### 3.5.3 The Action Stage.
During this stage, $V_{ego}$ should complete lane change smoothly to reduce the impact on traffic flow.

## 4. Detailed Calculations of the New DLCs Strategy

In this section, we list the distance calculations successively according to the interaction order.

### 4.1 Interaction between $V_{ego}$ and $V_{f1}$

The relationship between $V_{ego}$ and $V_{f1}$ in Fig. 5 is equivalent to the vehicle-following scenario. Keeping away from the forbidden area of $V_{ego}$ is the responsibility of $V_{f1}$.

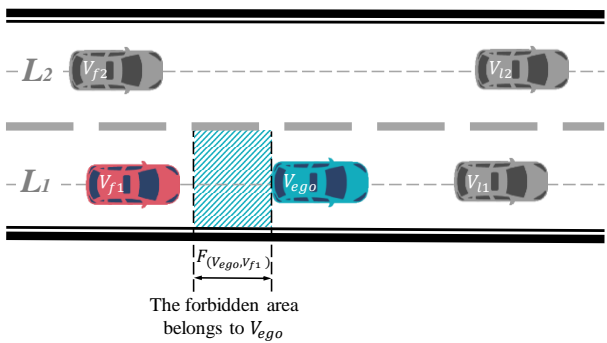

**Fig. 5.** *The assignment of right-of-way between $V_{ego}$ and $V_{f1}$.*

### 4.2 Interaction between $V_{ego}$ and $V_{l1}$

Similar to the above, this relation can be regarded as another vehicle-following scenario, as shown in Fig. 6.
The length of the forbidden area $F_{(V_{l1}, V_{ego})}$ can be expressed as

$$F_{(V_{l1}, V_{ego})} = \left[ v_{ego}\rho + \frac{v_{ego}^2}{2a_{ego,brake}} - \frac{v_{l1}^2}{2a_{max,brake}} \right]^+ \quad (4\text{-}1)$$

$$a_{ego,brake} = a_{min,brake} + \frac{v_{ego}}{v_{max}}\left(a_{max,brake} - a_{min,brake}\right) \quad (4\text{-}2)$$

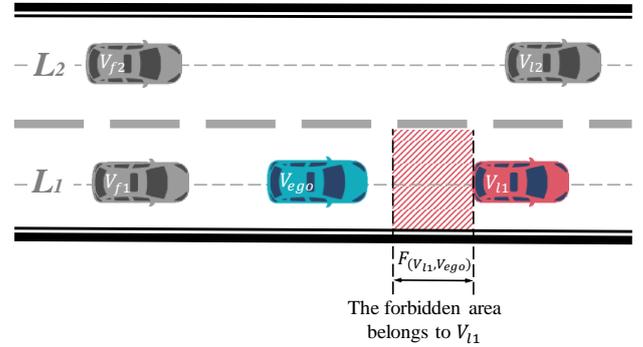

**Fig. 6.** *The assignment of right-of-way between $V_{ego}$ and $V_{l1}$.*

### 4.3 Interaction between $V_{ego}$ and $V_{l2}$

From the perspective of $V_{ego}$, $V_{l2}$ has a right-of-way advantage similar to $V_{l1}$, as shown in Fig. 7. Therefore, $F_{(V_{l2}, V_{ego})}$ can be expressed as

$$F_{(V_{l2}, V_{ego})} = \left[ v_{ego}\rho + \frac{v_{ego}^2}{2a_{ego,brake}} - \frac{v_{l2}^2}{2a_{max,brake}} \right]^+ \quad (5\text{-}1)$$

$$a_{ego,brake} = a_{min,brake} + \frac{v_{ego}}{v_{max}}\left(a_{max,brake} - a_{min,brake}\right) \quad (5\text{-}2)$$

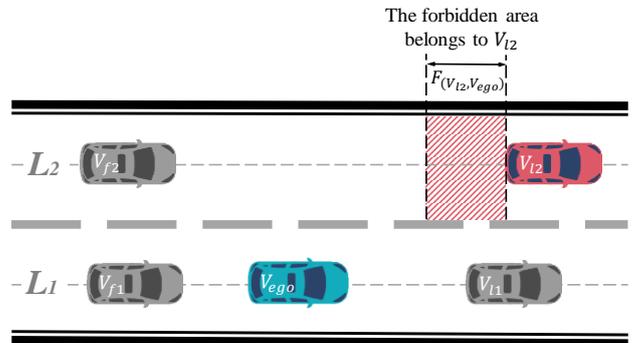

**Fig. 7.** *The assignment of right-of-way between $V_{ego}$ and $V_{l2}$.*

### 4.4 Interaction between $V_{ego}$ and $V_{f2}$

Different from the above three parts, the right-of-way assignment between $V_{ego}$ and $V_{f2}$ is more complicated because it may involve a request and response process. The front area of $V_{f2}$ can be divided into three areas, as shown in Fig. 8.

The length of $F_{(V_{f2}, V_{ego})}$ can be calculated by

$$F_{(V_{f2}, V_{ego})} = \left[ v_{f2}\rho_{human} + \frac{v_{f2}^2}{2a_{f2,brake}} - \frac{v_{ego}^2}{2a_{max,brake}} \right]^+ \quad (6\text{-}1)$$

$$a_{f2,brake} = a_{min,brake} + \frac{v_{f2}}{v_{max}}(a_{max,brake} - a_{min,brake}) \quad (6\text{-}2)$$

Since the reactive side of emergency braking is $V_{f2}$, it should be conservatively assumed to be a human driving vehicle with a longer reaction time $\rho_{human}$.

The ownership of the negotiable area will transfer with the different responses of $V_{f2}$, as shown in Fig. 8. The boundary of this area can be calculated as

$$F_{(V_{f2},V_{ego})} + N_{(V_{f2},V_{ego})} = \left[ v_{f2}\rho_{human} + \frac{a_{max,accel}\rho_{human}^2}{2} + \frac{(v_{f2} + a_{max,accel}\rho_{human})^2}{2a_{min,brake}} - \frac{v_{ego}^2}{2a_{max,brake}} \right]^+ \quad (7)$$

The meaning of this length is that *even when using minimum acceleration to brake, the following vehicle of the target lane can also avoid any impact caused by the merging vehicle*. Therefore, we use this as the criterion for the direct merging gap.

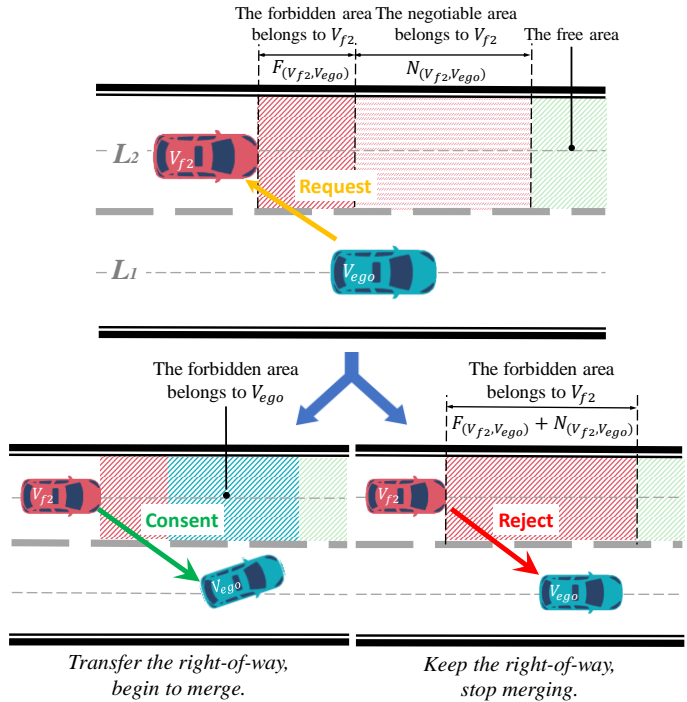

**Fig. 8.** The assignment of right-of-way between $V_{ego}$ and $V_{f2}$.

## 5. Results and Discussion on Efficiency

When assumptions A5 and A6 hold, both the RSS strategy and the new strategy can prevent collisions within the whole lane-changing process. Therefore, this section focuses on the comparison of their impact on traffic efficiency. We design two simulation experiments to observe the average time cost of DLC and the success rate of DLC in constant time under different strategies.

### 5.1 Simulation Settings

To closely approximate real traffic, the lane-changing scenario in the simulation system is constructed as shown in Fig. 2. We denote the speed of the ego as $v_{ego}$, and it plans to merge into the adjacent lane occupied by the fleet with average speed $v_{L2}$.

In reference [56], the researchers tried all the proposed distribution models to fit the empirical headway data and found that the log-normal model yielded the best fitting to real road conditions. Thus, we used the log-normal model to describe the headway of lane $L_2$, which is written as:

$$f(h) = \frac{1}{\sqrt{2\pi}\sigma h} \exp\left(-\frac{[\ln h - \mu]^2}{2\sigma^2}\right) \quad (8)$$

where variable $h$ denotes the possible value of the headway, $\mu$ is the location parameter and $\sigma$ is the scale parameter. The mathematical expectation of $h$ is:

$$\frac{3600}{\lambda} = E(h) = \exp(\mu + \sigma^2/2) \quad (9)$$

where $\lambda$ denotes the value of traffic flow. According to [56], we assume $\sigma = 0.8$ in this part and use Equation (9) to calculate parameter $\mu$ under different $\lambda$.

According to the SAE [57] and FMVSS135 [58] standards, we set the other major parameters required for the simulation as follows: $l_{vehicle} = 5\,\text{m}$, $a_{max,accel} = 2\,\text{m/s}^2$, $a_{min,brake} = 2\,\text{m/s}^2$, $a_{max,brake} = 6\,\text{m/s}^2$, $\rho = 0.1\,\text{s}$, $\rho_{human} = 1\,\text{s}$, $v_{max} = 30\,\text{m/s}$, $v_{ego} = 20\,\text{m/s}$ and $v_{L2} = 20\,\text{m/s}$.

Based on the above settings, we carried out the following two representative simulation experiments. It should be noted that we also tried other parameter settings, but these only caused insignificant numerical differences, which did not affect the conclusions in this paper.

### 5.2 The Average Time Cost of DLC

In the first experiment, we recorded the time cost for lane changes under different traffic flows ranging from 200 veh/h to 1600 veh/h. To reduce any errors caused by chance, we carried out 10000 tests and took the average.

The simulation results are shown in Fig. 9. We highlight the different time periods with three colors. The results illustrated that with the increase in $\lambda$, the required time of the RSS strategy increases rapidly. When $\lambda > 1160\,\text{veh/h}$, the time cost of more than 2 minutes becomes unbearable. Conversely, the new strategy is not significantly affected by the increase in $\lambda$. This is partly because the new strategy has a shorter min-acceptable gap; more importantly, the existence of negotiation makes it easy for $V_{ego}$ to find an opportunity to change lanes. Based on the new strategy, the merging maneuvers can be completed in 1 minute if more than 25% of vehicles consent to the request. More importantly, the existence of negotiation makes it easy for $V_{ego}$ to find an opportunity to change lanes. Based on the new strategy, the merging maneuvers can be completed in 1 minute if more than 25% of vehicles consent to the request.

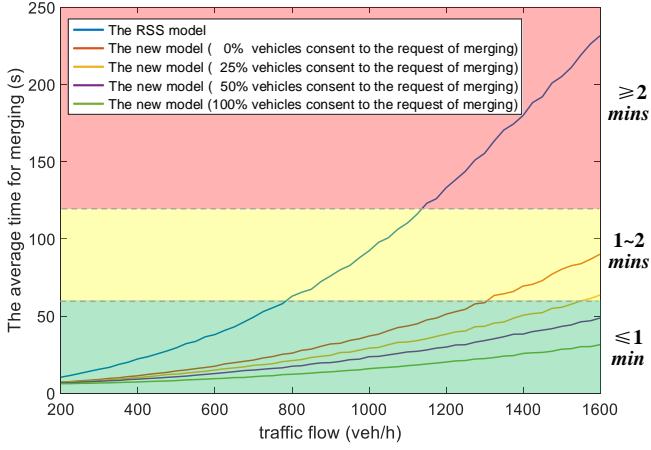

***Fig. 9.*** *The average time cost under different traffic flows.*

### 5.3 The Success Rates of DLC

In the second experiment, we checked the success rates for lane changes at constant times ranging from 20 s to 180 s. In addition, we carried out experiments under $\lambda = 600\,\text{veh/h}$ and $\lambda = 1200\,\text{veh/h}$. Similar to the first experiment, we carried out 10000 tests and took the average.

The results are shown in Fig. 10. Similar to the conclusion of the above experiment, it can be seen that the effect of the RSS strategy is not ideal under high traffic flows. As a comparison, the success rate of the new strategy under the same conditions is stable above 69%. Furthermore, the success rate of the new strategy is significantly affected by the intention of the vehicles from the target lane, whose high tolerance of requests can effectively reduce the waiting time of $V_{\text{ego}}$. Overall, the success rate of the new strategy is greatly improved compared with RSS, which benefits from the negotiation mechanism and shorter min-acceptable gap.

## 6. Results and Discussion on Safety

In above section, the efficiency comparison results are built on the basis that assumptions A5 and A6 (from *Chapter 3.1*) hold. Since the uncertainty is removed, the whole simulation experiment is controllable. Therefore, to validate the model safety, it is also necessary to discuss situations when these two assumptions fail. In this section, we discuss multivehicle lane-changing scenarios and negotiation failure scenarios and perform a safety testing experiment under these extreme cases in *Chapter 6.3*.

### 6.1 Multivehicle Lane-changing Scenarios

The above content focuses on the two-lane scenarios introduced in *Chapter 2.2*, where $V_{\text{ego}}$ is the only vehicle with the intention to change lanes. However, this assumption does not always hold on real roads. During the longitudinal position adjustment stage, the merging vehicle sometimes has to face interference caused by merging maneuvers from other vehicles. When two vehicles are planning to merge simultaneously, rational human drivers usually determine the right-of-way according to the FCFS principle. In more extreme cases, such as when three or more vehicles are merging together, giving up merging temporarily for safety is a more common choice.

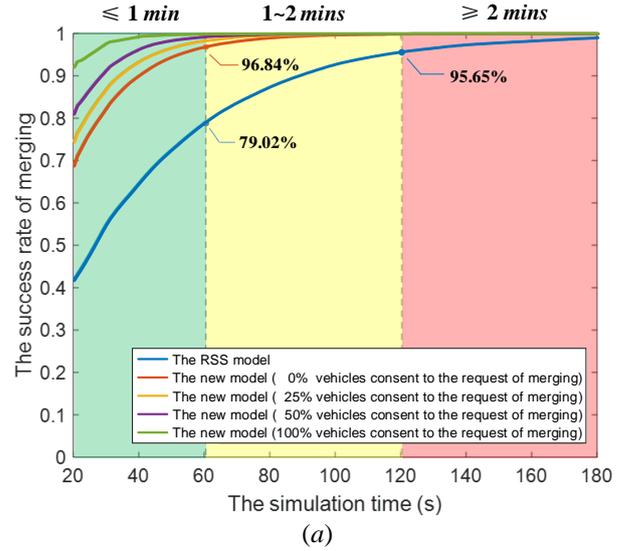

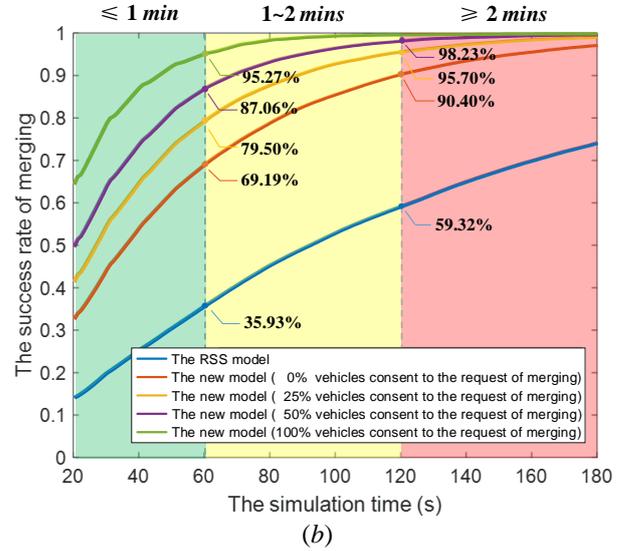

***Fig. 10.*** *The success rate of lane changes in constant time.*
(*a*) $\lambda = 600\,\text{veh/h}$ ; (*b*) $\lambda = 1200\,\text{veh/h}$ .

Therefore, it is necessary to determine how AVs should react when two vehicles show a desire to change lanes. We expand the original scenario to the four lanes shown in Fig. 11 and add two vehicles $V_a$ (in $L_0$) and $V_b$ (in $L_3$) as new research objects. Moreover, we summarize the merging behaviors that may cross the trajectory of $V_{\text{ego}}$ into seven types, marked ① to ⑦ in Fig. 11. When addressing these scenarios, the right-of-way rules can be introduced similarly to clarify the relationship between the merging initiators and $V_{\text{ego}}$. According to the relative strength of the right of way, we can categorize these seven types into the following three situations:

*6.1.1 Ego vehicle has a higher right-of-way*: from ① to ⑤. In this case, $V_{\text{ego}}$ has the right to choose between continuing the merge or giving up, and the surrounding vehicles have obligations to cooperate with this choice. Among them, the initiators of maneuvers ①② and ③ should respect the right-of-way area behind $V_{ego}$ or bear responsibility for possible accidents. Unlike ①② and ③, maneuvers ④ and ⑤ may invade the negotiable area of $V_{ego}$, so those maneuvers can only be executed after acquiring the consent of $V_{ego}$.

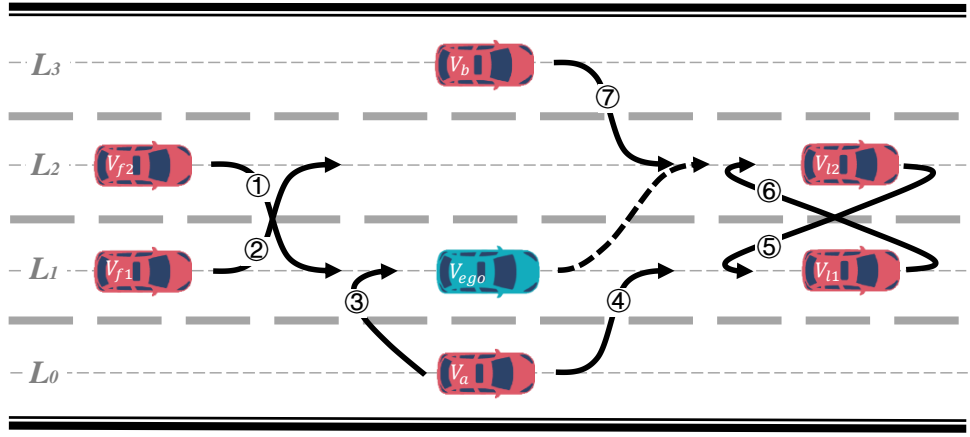

**Fig. 11.** *Seven multivehicle merging scenarios that may affect $V_{ego}$.*

*6.1.2 Ego vehicle has a lower right-of-way:* ⑥. In this case, $V_{ego}$ should actively yield to avoid entering the forbidden area of $V_{l1}$. After the action of $V_{l1}$, if the remaining distance ahead is insufficient to accommodate another merge, $V_{ego}$ should quickly abandon and readjust its longitudinal position to find a new opportunity.

*6.1.3 The right-of-way relationship cannot be determined:* ⑦. This means both $V_b$ and $V_{ego}$ plan to enter the same area of $L_2$, but neither has an obvious advantage in terms of right of way. According to the FCFS principle, the vehicle first showing a merging intention ought to acquire the right-of-way advantage, and the other vehicle should respect this assignment result. To increase reliability and integrity, the failure cases against this rule are discussed individually in the next section.

### 6.2 Negotiation Failure Scenarios

In a mixed traffic flow, understanding errors (of human-driving vehicles) or algorithm errors (of AVs) may cause the vehicles surrounding $V_{ego}$ to refuse to negotiate or refuse to accept the negotiated results. We call this situation the "negotiation failure," which is specifically manifested as a vehicle suddenly entering the forbidden area of $V_{ego}$ without warning during its lane-changing action stage. Because this maneuver is unexpected, it is necessary to develop targeted measures; otherwise, the risk of accidents will increase greatly. Fig. 12 shows two typical negotiation failure scenarios.

*6.2.1 Wrong acceleration of $V_{f2}$.*

i) If when the acceleration of $V_{f2}$ is detected, $V_{ego}$ does not enter $L_2$. It should stop merging immediately.

ii) If when the acceleration of $V_{f2}$ is detected, $V_{ego}$ has partially entered $L_2$ and space ahead is sufficient. It can accelerate like $V_{f2}$ and complete merging.

iii) If when the acceleration of $V_{f2}$ is detected, $V_{ego}$ has partially entered $L_2$ and space ahead is not sufficient. ego should stop merging and return to $L_1$ immediately. Assuming that $V_{f2}$ accelerates after $V_{ego}$ starts merging $t_{merging}$, the time $t_{elude}$ left for $V_{ego}$ to avoid collision and return to $L_2$ can be calculated as follows:

$$t_{elude}=\sqrt{\left[(V_{ego}-V_{f2})t_{merging}+F_{(V_{f2},V_{ego})}\right]/a_{accel}}-\rho \quad (10)$$

After bringing in common parameters, $t_{elude} > t_{merging}$ can be obtained by simulation calculation, which proves that $t_{elude}$ is sufficient for $V_{ego}$ to perform a reverse maneuver.

*6.2.2 Wrong merging of $V_b$.*

i) If when the merging of $V_b$ is detected, $V_{ego}$ has not yet entered $L_2$. $V_{ego}$ should stop merging immediately.

ii) If when the merging of $V_b$ is detected, $V_{ego}$ has partially entered $L_2$. $V_{ego}$ should give up merging and return to $L_1$ with the largest possible lateral acceleration. If the lateral speed of $V_b$ is excessive or the original space of $L_1$ is occupied, $V_b$ obviously needs to bear full responsibility for the potential collision.

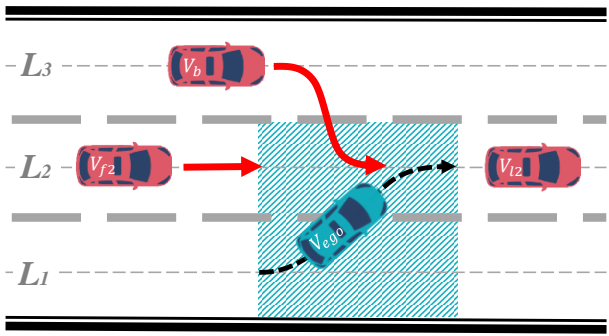

**Fig. 12.** *Two typical negotiation failure scenarios.*

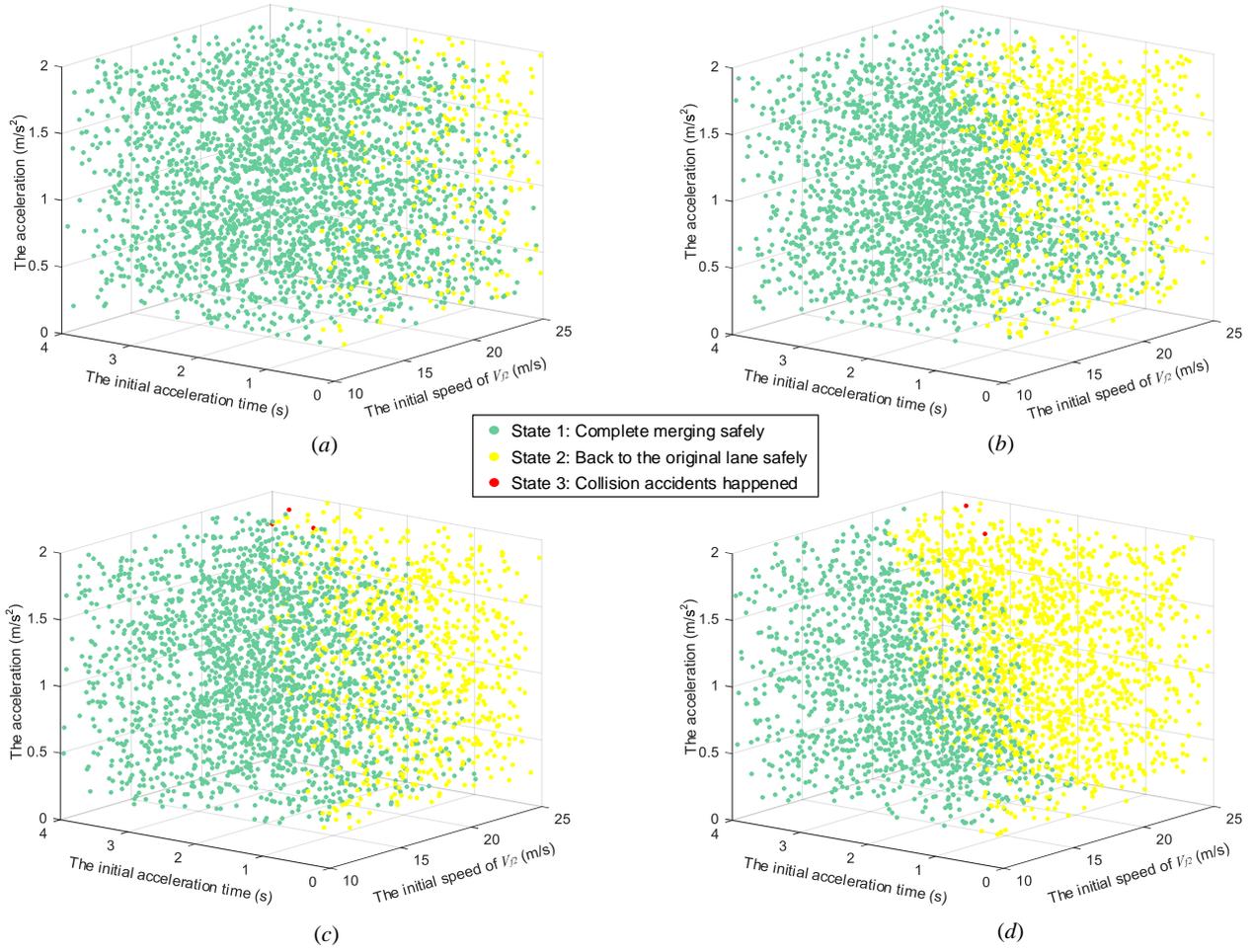

***Fig. 13.*** *The experimental results of safety simulation for two strategies in the extreme cases of negotiation failure.*
(*a*) *The new strategy,* $\lambda = 600 \text{ veh/h}$; (*b*) *The RSS strategy,* $\lambda = 600 \text{ veh/h}$;
(*c*) *The new strategy,* $\lambda = 1200 \text{ veh/h}$; (*d*) *The RSS strategy,* $\lambda = 1200 \text{ veh/h}$.

### 6.3 Numerical Testing Results

To verify the safety in extreme cases, we perform multiple simulation experiments for negotiation failure. Similar to the last simulation, we use the same traffic scenarios and traffic flow models as in *Section 5*. On this basis, we assume that $V_{ego}$ request for merging to the negotiation area of $V_{f2}$ and receive a acquiescence, but the negotiation failure will occur in each DLC experiment, i.e., $V_{f2}$ will suddenly accelerate to interrupt the merging process until reaching the maximum lane speed limit (30 m/s).

In the experiment, we repeat the lane change experiment 10,000 times at different initial speeds of $V_{f2}$ (from 10 m/s to 25 m/s), different initial acceleration times (from 0 s to 4 s), different accelerations (from 0 to 2 m/s²), and different traffic densities (600 veh/h and 1200 veh/h) to observe the safety performance of the new strategy and the RSS strategy. According to *Chapter 6.2.1*, when this unexpected acceleration behavior is detected, the AV faces three possibilities. State 1: Complete the merging safely if the distance ahead is sufficient. State 2: Drive back to the original lane if it has not yet entered $L_2$ or the distance is insufficient. State 3: a collision will occur if safe avoidance is not available.

***Table 2.*** *The distribution of experimental results in the extreme cases of negotiation failure.*

| Traffic flow $\lambda$ | Merging strategies | The proportion of results | | |
|---|---|---|---|---|
| | | State 1 | State 2 | State 3 |
| 600veh/h | New | 92.28% | 7.72% | 0 |
| | RSS | 72.14% | 27.86% | 0 |
| 1200veh/h | New | 73.38% | 26.58% | 0.04% |
| | RSS | 54.32% | 45.62% | 0.06% |

Finally, the experimental results are displayed in Fig. 13, and the distribution of the results is summarized in Table 2.

From the results, we can see that the AVs with the RSS strategy returns to the original lane in more cases because it has a longer safety distance and thus a stricter determination of potential dangers. While AVs with the new strategy would choose to complete lane changes in more cases, their difference in safety performance is not significant.

At $\lambda = 600 \text{ veh/h}$, both strategies can ensure 100% safety. Focusing on the collision cases at $\lambda = 1200 \text{ veh/h}$, we find that there was a small chance for $V_{ego}$ of falling into inescapable safety dilemmas under both strategies, which leads to all the crashes in the results. As shown in Fig. 14, if $V_{f2}$ incorrectly accelerates and $V_{f1}$ mistakenly believes that $V_\text{ego}$ has already changed lanes and accelerates early, the collision will be inevitable. Under this circumstance, the appropriate action is to minimize damage by choosing the correct collision object. Undoubtedly, the vehicle violating the right-of-way rules first should be primarily responsible. However, in actual traffic, the division of responsibilities for those cases still requires further and careful discussion between relevant traffic lawmaking agencies and autonomous driving technology providers.

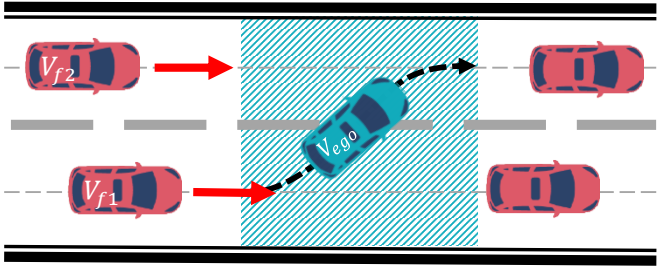

*Fig. 14.* A dilemma caused by the refusal of multiple vehicles to implement the consensus results.

The primary purpose of this work is to develop a communication-based lane-changing strategy under the framework of rational behaviors. Thus, the new strategy achieves better traffic efficiency performance with an approximate safety of RSS, which is sufficient to prove its superiority and practical value.

## 7. Conclusions

In this paper, we propose a communication-based right-of-way assignment strategy to improve the lane-changing performance of the RSS. Our main work is decomposing the DLC process into three stages and translating the collision avoidance conditions of each stage into a formulized expression. Uncomplicated calculations and simulations show that the introduction of negotiation can improve the utilization of limited road resources, thus ensuring that the right-of-way assignment is more efficient and reasonable.

We believe these proposed right-of-way principles are universal and applicable, which means that more verifications and improvements are needed [59]. Currently, we are applying this communication-based formulized method to more scenarios, such as ramp scenarios and non-signal intersections [53]. We believe that under the premise of ensuring safety, negotiated driving in a human-understandable manner will be an essential task for encouraging the use of AVs in mixed traffic [38].

### Acknowledgments

This work was supported by the National Natural Science Foundation of China (61790565)